\documentclass[sigconf]{acmart}

\usepackage{cite}
\usepackage{algorithmic}
\usepackage{graphicx}
\usepackage{textcomp}
\usepackage{xcolor}
\usepackage{lipsum}
\usepackage{hyperref}
\usepackage{multirow}
\usepackage{comment}
\usepackage{cleveref}
\usepackage{tabularx}
\usepackage{subcaption}
\usepackage{caption}
\usepackage{enumitem}
\usepackage{multicol}

\copyrightyear{2024}
\acmYear{2024}
\setcopyright{acmlicensed}\acmConference[EuroUSEC 2024]{The 2024 European Symposium on Usable Security}{September 30-October 1, 2024}{Karlstad, Sweden}
\acmBooktitle{The 2024 European Symposium on Usable Security (EuroUSEC 2024), September 30-October 1, 2024, Karlstad, Sweden}
\acmDOI{10.1145/3688459.3688463}
\acmISBN{979-8-4007-1796-3/24/09}

\begin{document}
	
\title{From Chaos to Consistency: The Role of CSAF in Streamlining Security Advisories}
\subtitle{Authors' version; to appear in the Proceedings of the 2024 European Symposium on Usable Security (EuroUSEC 2024)}

\author{Julia Wunder}
\affiliation{
    \institution{IT Security Infrastructures Lab, Friedrich-Alexander Universität Erlangen-Nürnberg (FAU)}
    \city{Erlangen}
    \country{Germany}
}
\email{julia.wunder@fau.de}

\author{Janik Aurich}
\affiliation{
    \institution{IT Security Infrastructures Lab, Friedrich-Alexander Universität Erlangen-Nürnberg (FAU)}
    \city{Erlangen}
    \country{Germany}
}
\email{janik.aurich@fau.de}

\author{Zinaida Benenson}
\affiliation{
    \institution{IT Security Infrastructures Lab, Friedrich-Alexander Universität Erlangen-Nürnberg (FAU)}
    \city{Erlangen}
    \country{Germany}
}
\email{zinaida.benenson@fau.de}

\renewcommand{\shortauthors}{Wunder et al.}
 
\begin{abstract}
Security advisories have become an important part of vulnerability management. They can be used to gather and 
distribute valuable information about vulnerabilities. Although there is a predefined broad format for advisories, it is not really standardized. As a result, their content and form vary greatly depending on the vendor. Thus, it is cumbersome and resource-intensive for security analysts to extract the relevant information. 
The Common Security Advisory Format (CSAF) aims to bring security advisories into a standardized format which is intended to solve existing problems and to enable automated processing of the advisories. However, a new standard only makes sense if it can benefit users. Hence the questions arise: Do security advisories cause issues in their current state? Which of these issues is CSAF able to resolve? What is the current state of automation? 

To investigate these questions, we interviewed three security experts, and then conducted an online survey with 197 participants. The results show that problems exist and can often be traced back to confusing and inconsistent structures and formats. CSAF attempts to solve precisely these problems. However, our results show that CSAF is currently rarely used. Although users perceive automation as necessary to improve the processing of security advisories, 
many are at the same time skeptical. One of the main reasons is that systems are not yet designed for automation and a migration would require vast amounts of resources. 
\end{abstract}



\maketitle

\section{Introduction}
\label{sec:introduction}

The IT sector is growing steadily from year to year, and the number of connected devices is increasing. By 2030, it is estimated that more than 29 billion devices will be connected \citep{statista-iot}. However, the complexity increases as well, as many systems build and depend upon each other. No system is perfect and thus, the number of vulnerabilities also increases every year. This creates major challenges for companies, as they need to maintain an overview of vulnerabilities and affected systems in order to decide which vulnerabilities need to be addressed immediately which can wait or do not need to be addressed at all. Prioritization is important as vulnerability management resources are often limited.

Security advisories are a popular approach 
to provide this prioritization overview \citep{2021-miranda}. They contain information on vulnerabilities and affected products, and provide guidance on how to deal with the vulnerability. They are usually provided by vendors for their products and distributed by aggregators such as the NVD\footnote{National Vulnerability Database, \url{https://nvd.nist.gov/}}.
However, the security advisories are usually not available in a predefined format and the vendors themselves often decide how their security advisories are presented and distributed. Companies relying on solutions from various vendors often find themselves caught in a long chain of dependencies. 
The literature has shown that installing software updates is very time-consuming and often causes problems \citep{2019-li, 2020-tiefenau}, as each company uses various systems that use distinct products from different vendors, creating a complex and difficult-to-maintain environment. The challenge for a security analyst is to reliably find, filter and process information on software vulnerabilities in their organization. However, it is not known exactly how this process is carried out, what role security advisories play in it and what hurdles can be encountered.
There are indications that the current state of security advisories is far from ideal. For instance, obtaining relevant information is often tedious due to the many irrelevant security advisories that are received \citep{2021-miranda, 2020-farhang}. Sometimes essential information is also missing from an advisory, making it hard to fully understand the impact of the affected system to make an appropriate decision \citep{2018-kula}. This implies that a standardized format with a uniform structure is needed.

The Common Security Advisory Format\footnote{\url{https://docs.oasis-open.org/csaf/csaf/v2.0/csaf-v2.0.html}} (CSAF) is designed to enable automated processing of security advisories due to its standardized format. Nevertheless, a new standard will only be accepted by users if it solves existing problems and can easily be adopted. In order to draw a comparison as to whether CSAF can lead to an improvement, the current situation must be investigated.
This raises the questions about how security advisories are currently processed, what problems can arise and whether CSAF provides a framework that addresses the existing challenges.

\paragraph{Contributions}
\label{sec:contributions}

We summarize the contributions of this work as follows:
\begin{itemize}
    \item We shed light on how security advisories are currently received, processed and what influences the decision-making process. For example, advisories are usually received through one main channel, such as email. In addition, multiple mailing lists are often subscribed simultaneously due to the concern of missing relevant advisories. How a decision is finally made with the advisories depends heavily on the company policy, although the probability and potential impact of a vulnerability are consistently considered important. 
    \item We highlight problems caused by the inconsistent format of security advisories. For example, inconsistent product identifiers defined by vendors complicate the process of identifying whether the advisory is relevant, as this information often has to be extracted manually. Thus, automatic pre-filtering is made more challenging, and non-relevant advisories are often received.
    \item We show that there is a great need and desire for a standardized format and automated processing. However, some users do not yet use automation, as adapting their systems to it is too resource-intensive.
    \item We show that CSAF can improve the processing of security advisories, as the machine-readable, standardized format enables automatic processing and saves resources. Nevertheless, solving some problems remains in the hands of the vendors, such as the completeness of the information contained in the security advisories. CSAF specifies a format, but not all entries in this format are mandatory.
\end{itemize}

\paragraph{Outline}
\label{sec:outline}

The paper is structured as follows. First, in \Cref{sec:background}, we give insight into the background of security advisories and CSAF, outline related work and present the research questions. Next, in \Cref{sec:prelim-study}, we describe the qualitative preliminary study in which we conducted interviews. The design and results of the quantitative main survey is described in \Cref{sec:main-study}. The findings are then discussed in \Cref{sec:discussion} and connected to related work and the real world. This work is concluded in \Cref{sec:conclusion}, where we provide a brief summary and outlook for future research.

\section{Background}
\label{sec:background}

\subsection{Security Advisories}
\label{sec:security-advisories}
Security advisories are documents that are usually issued by vendors consisting of valuable information about a vulnerability in one or more of their products. Sometimes, advisories are also issued by independent security researchers or cyber security organizations on a vulnerability they have discovered or researched. The main goal of these advisories is to inform users of the product, which are usually organizations, as well as other security researchers about these vulnerabilities in a comprehensive manner and also give recommendations on how to deal with them. This makes security advisories a very important tool for combating security threats which is why they are widely used today.
As there was previously no recognized standard for them, security advisories were and still mostly are shared in various formats that range from simple text files to proprietary data types. Their content as well as its ways of dissemination are mostly dependent on the issuer. Still, they generally follow the same content structure which consists of a list of affected products, followed by details about the vulnerability and a recommendation for action. However, these sections differ a lot in detail and writing quality, depending on the author. 

As far as the distribution of these documents is concerned, a variety of different methods in various combinations are applied. These include, but are not limited to, social media posts, blogs or mailing lists. Often users have to actively take care of obtaining security advisories. Some platforms or tools by different vendors also have made it their mission to simplify this process, by offering a service that collects, filters and distributes security advisories from various sources to interested parties. 

\subsection{CSAF}
\label{sec:csaf}
The Common Security Advisory Framework (CSAF) consists of a specification on how security advisory documents in the CSAF format are to be structured and which information they can contain as well as a set of tools that facilitate the creation, processing and distribution of them. The framework is created and maintained by OASIS\footnote{Organization for the Advancement of Structured Information Standards, \url{https://www.oasis-open.org/}}, a well-known non-profit organization committed to developing open standards and aims to be the de facto globally known and used standard for generating security advisories. To achieve this, OASIS has the support of numerous large international institutions and corporations, such as Cisco, Oracle or Siemens, which were actively involved in the development process.
CSAF documents are written in JSON\footnote{JavaScript Object Notation, \url{https://www.json.org/json-en.html}} and are generally divided in three main sections.
The document section contains various metadata of the document itself, such as the title, category and publisher details. The product tree section lists every single product that is referenced in the advisory as well as their relation to other products. It additionally can contain unique identifiers or identification helpers for them. The vulnerabilities section describes one or multiple vulnerabilities of the aforementioned products in great detail and also provides the reader with instructions on dealing with them. This is often accompanied by references to known vulnerability scores or descriptors, such as CWE\footnote{Common Weakness Enumeration, \url{https://cwe.mitre.org/}}, CVE\footnote{Common Vulnerabilities and Exposures, \url{https://www.cve.org/}} or CVSS\footnote{Common Vulnerability Scoring System, \url{https://www.first.org/cvss/}}.
Only a subset of the provided fields are mandatory for a CSAF document to be valid and a lot are optional.
OASIS also thought about the distribution of CSAF documents and therefore put an infrastructure of issuers and distributors of CSAF documents in place which allows users of the CSAF standard to quickly aggregate and filter advisories from multiple sources, depending on their needs.

\subsection{Related Work}
\label{sec:related-work}
As CSAF is very new, there exists no current research on its possible effect on vendors and users. Security advisories are also not a particularly well-studied topic, although some work has examined them in different contexts. User studies about the update behavior of system administrators and developers proved to be very useful for our studies, since they  investigated the handling of security advisories to a certain extent, even if these were not the focal point.

\subsubsection{Automation}

\citet{2008-fenz-semantic} analyzed structures of existing security advisories with the aim of identifying standardized semantics to enable automated processing of advisories. The authors evaluated the advisories primarily in terms of semantic usefulness, information complexity and distribution. The results indicate that none of the existing formats met the authors' criteria. The paper from 2008 shows how long the desire for a standardized format has been under discussion. Building on this work, \citet{2008-fenz-fortification} developed a framework that converts security advisories from different sources into a machine-readable format to enable automation.
\citet{2017-ramnani} focused on the format of security advisories and their automated processing. They used pattern recognition and natural language processing techniques to extract valuable information from large quantities of unstructured or semi-structured vulnerability information. For this purpose, they developed a prototype that they then evaluated on a test set. The authors emphasize the importance of automatic processing of security advisories.

\subsubsection{Processing of Security Advisories and Updates} 
\citet{2019-li} conducted a quantitative and qualitative study on how often and when system administrators perform updates. The authors identified five phases for update processing: 1) the system administrator is informed about the update and begins to gather information about it; 2) a decision is then made as to whether the update should be carried out or not; 3) the systems are prepared and the update is carried out, whereby 4) the system administrator communicates the update to the employees choosing a suitable time frame; 5) any problems that the update may have caused are resolved. The results indicate that the phases are quite complex and time-consuming for the system administrators. According to the authors, many of the steps could be optimized using automation.
The same topic was addressed by \citet{2020-tiefenau}, who also interviewed and surveyed system administrators. They examined the update process and identified similar phases: information gathering, decision, update preparation, test run, post-installation.
These first two steps -- information gathering and decision-making -- in the context of security related updates revolve primarily around security advisories and are of particular importance to our studies. 
Considering previous research on automation and the emphasis of CSAF on automated processing, we think that processing of gathered information (including security advisories) represents a crucial additional step between information gathering and decision-making. Using the insights of related work \citep{2019-li, 2020-tiefenau}, the processing of security advisories can be classified into three main steps: 1) information gathering, 2) information processing and 3) decision-making. We used these steps to structure the interview guide and the survey questionnaire. The steps are briefly explained in the following.

\paragraph{Information gathering} This step involves determining which channels are used and how the information reaches the user. \citet{2021-miranda} analyzed security advisory platforms to determine how vulnerabilities are first published and distributed to other platforms. They found two types of platforms: information sources and aggregators. Vulnerability information spreads through a network of connected platforms. To investigate if developers keep software libraries up to date, \citet{2018-kula} examined various GitHub projects and found many outdated dependencies which contained security vulnerabilities. When inquired about this issue, developers often replied that they were not aware of the outdated dependencies. Most of them stated that they simply do not have the time to search for corresponding security advisories and generally lack resources to update their dependencies. \citet{2020-farhang} analyzed various bulletins which vendors often publish for their own platforms, such as Android, to summarize security-relevant events. The authors show that the vendors often do not use the standardized CVEs, but their own identifiers in the bulletins which makes it unclear whether all critical vulnerabilities are named and whether the information is relevant. These studies show that a main channel to receive advisories would be desirable, as this would make the process less complex. This demonstrates that it is highly complex for security analysts to receive all relevant information. Often, all sorts of security advisories are received, some of which are not relevant \citep{2021-miranda}. Another problem is that the information is often distributed across different sources and users have to tediously gather and filter it \citep{2021-miranda, 2018-kula}. Trustworthiness of the sources also plays a role \citep{2020-jenkins}, which CSAF ingrained in their design, as CSAF trusted providers have to sign and hash their advisories \citep{csaf-trust}.

\paragraph{Information processing}
Once information has been collected, it is processed either manually or with the help of tools \citep{2021-miranda}. If tools are used, it is particularly important that the information is available in a complete and machine-readable format, as otherwise it cannot be processed \citep{2008-fenz-semantic, 2008-fenz-fortification}. The absence of a standardized format means that the information often appears redundantly in the security advisories or is widely distributed across different sections, such that filtering takes a lot of time. Users would therefore like to have better options for filtering the information with the help of a tool \citep{2021-miranda}.

\paragraph{Decision making} The final step is to decide how to deal with the vulnerability. It is particularly important that the vulnerability is described in as much detail as possible in the security advisories to fully understand its impact on the affected systems \citep{2018-kula}. Users would like to have more opportunities to contact other affected users or the vendors directly in order to simplify the decision-making process \citep{2018-kula}. The exact procedure to be followed afterwards depends on company policies and whether a patch would threaten the current availability of systems or resources \citep{2019-li, 2020-tiefenau}.

\subsection{Research Questions}
\label{sec:research-questions}

Security advisories play an important role in vulnerability management. Therefore, it is important to know how exactly users work with security advisories and which steps they go through when processing them. For example, from which sources they receive advisories, what information is relevant and which actions are taken.

The literature as described in \Cref{sec:related-work} indicates that the processing of security advisories is very time-consuming and complex. Possible problems may arise at various points in the processing workflow and further tie up valuable resources. We thus investigate which problems may occur based on the current state of security advisories.

An important feature of CSAF is the ability to process security advisories automatically. However, the precondition for this is that users are willing to use automation and to adapt their own systems and setups if necessary. We shed light on the question of how automation is currently used for security advisories and what the users' intentions are, for example, whether they are already using automation or are planning to introduce it soon.

One finding from the literature was that problems often arise due to the inconsistent structure of advisories. CSAF attempts to tackle this problem by creating a standardized and machine-readable format that can be processed automatically. This raises the question to what extent CSAF is in current use, and which other problems CSAF plans to solve with its current approach.

To summarize, we consider the following research questions:

\begin{enumerate}[itemindent=2em,labelsep=0.5em,label=RQ\arabic*:]
    \item How are security advisories currently processed?
    \item Which issues do security advisories cause in their current state?
    \item What is the current state of automation for processing security advisories?
    \item Which security advisory issues can CSAF address and which not?
\end{enumerate}

The research questions are investigated using a qualitative preliminary study and a quantitative main survey. RQ1-RQ3 are examined directly by the results in \Cref{sec:results}. RQ4 is evaluated by the interpretation of the results in \Cref{sec:discussion}.

\section{Preliminary Study - Interviews}
\label{sec:prelim-study}

\subsection{Study Design}
\paragraph{Interview Guide and Testing} Three qualitative interviews were conducted to identify findings that would be further investigated in the quantitative survey. The interview guide was developed from similar topics previously found in the literature review. The aim of the preliminary study was to discuss the problems that were identified during literature review in more detail. This was done to get a rough estimation of how widespread and severe they are perceived, and to add other aspects that were not previously considered, such as the fear of missing relevant security advisories.

The interview guide can be found in \Cref{sec:appendix-interview-guide}. First, the interview participants were asked to introduce themselves and give a brief overview of their workplace and work experience, for example how often they come into contact with security advisories. Next, they were queried about collecting information, for example from which sources they obtain advisories, whether they receive too many irrelevant advisories or if there are any other problems. This was followed by questions about the processing of security advisories, such as whether they are processed manually or automatically. We requested the participants to name the advantages and disadvantages of automation or manual work that is currently used in their day-to-day work. We then went on to discuss the decision-making process once the information from the security advisories had been processed. Among other things, they were asked which factors can influence a decision. Finally, the interviewee's demographic data was recorded and room was provided for further comments. To ensure that the questions were comprehensible and to estimate the duration of the interviews, a test interview with a colleague from our lab was conducted.

\paragraph{Ethics} The data protection office of the Friedrich-Alexander-Universität
Erlangen-Nürnberg (FAU) approved the study. Before the recording started, the participant was provided with an informed consent form that informed them about their rights of information and deletion, and the purpose of the study. There was no compensation and the participants could interrupt the interview at any time and decline to answer. During and after the interview, it was possible to let us know if certain information should not appear in the final transcript. The answers were pseudonymized and the recordings stored on a server of our university. All interviews were conducted remotely via Zoom.

\paragraph{Recruitment and Data Analysis} We focused on recruiting participants who work intensively with security advisories on a day-to-day basis and are familiar with CSAF to inquire about CSAF-specific topics, such as automation. To advertise the interviews and the subsequent survey, we gave a short talk about the project at a conference for security professionals in Germany\footnote{\url{https://www.dfn.de/event/30-dfn-konferenz-sicherheit/}} . This way, we recruited two interview participants and a third participant was recruited through personal contacts from our lab. The interviews took 22 minutes on average. They were recorded and then transcribed manually by one researcher. All personal references were removed in the process.
After that, the transcript was analyzed by one researcher in order to fill previously identified themes of security advisory processing from literature with content. For example, the exact procedure and possible problems were collected for each processing step. Due to the small sample size and the fact that the broad categories (information gathering, information processing, decision-making) had already been determined after the literature review, it was decided not to use coding. 
While reviewing the transcripts, new themes that had not previously appeared in the literature were added to the list of themes, such as the willingness for automation from a user perspective. 
After the interviews were analyzed, the identified themes were discussed by two researchers to draw conclusions.

\paragraph{Participants}
All three participants come from Germany, are male, have a university degree and were on average around 55 years old at the time of the interview. One of the participants works as head of IT, one is a system administrator. The third participant works as a service manager at a company that publishes security advisories. All participants have many years of professional experience with security advisories and vulnerability management.

\subsection{Results}

\paragraph{Processing of Security Advisories} To receive security advisories, the interviewees use the DFN\footnote{Deutsches Forschungsnetz (German research network), \url{https://www.dfn-cert.de/}} service which informs subscribers by email about vulnerabilities that affect them. Subscribers are also informed when something has changed in the vulnerability which the interviewees emphasized as particularly positive. However, one participant noted that he was unable to change the settings of his subscription and consequently also receives some security advisories that do not apply to his systems. The other participants also said that they occasionally receive security advisories that do not match their current system or setup.

One problem stated by the interviewees is a mismatch in the product identifiers which normally mark products with a unique code. 
According to the participants, there are often several different product identifiers for the same product, as companies assign their own identifiers internally. It is thus possible that various product identifiers are specified, but they refer to the same product.
It is then necessary to manually check which products are actually affected. This is described by the participants as very time-consuming, as the list of affected systems is often very long and in an inconsistent format, usually consisting only of a sequence of letters and numbers. This makes it particularly difficult to determine whether one's own system is affected by the vulnerability or not. P3 summarized this problem as: \emph{``If we write in S7-1511 and the customer wrote Sematic S7-1500 Family or something like that, [...] even if he has an asset management system and we have machine-readable advisories, the matching is — you could try it with AI or something — [...] not that simple. In other words, the issue of unique product identifiers is still unresolved throughout the industry.''} 
The interviewees consider automation to be difficult without a standardized format for product identifiers, as they fear that they will miss important advisories.

As described in \Cref{sec:security-advisories}, security advisories are usually provided by the manufacturers as PDFs or other text documents. 
According to the participants, another problem is the varying length and quality of the security advisories. For example, some advisories are too short, 
making it necessary to find additional sources. One participant noted that a certain manufacturer only offers links in its security advisories. 
Thus, the participants prefer the security advisories to be more detailed. 
In P1 opinion, \emph{``there can't really be too much [information], because if a document is well-structured, it's not a problem''}.

The participants see a standardized format for security advisories as very positive. According to the participants, the way the vulnerability is subsequently handled depends on its severity and internal company policies. For instance, whether it requires immediate fixing or whether a patch cycle should be waited for.

\paragraph{CSAF} P1 and P2 are familiar with CSAF, but do not currently use it. P3 is an employee of a company that has cooperated with OASIS 
to publish security advisories in the CSAF format. Currently, the security advisories are already processed semi-automatically according to the participants by pre-filtering them depending on the affected system. Although the participants stated some problems, such as mismatches in product identifiers and the cumbersome manual research of which systems are exactly affected, 
the interviewees do not see a significant need for further automation. 
P3, who works as a service manager and therefore deals with customers who process security advisories on a daily basis, also emphasized that many users are currently not ready for automation. This is because it requires a functioning asset management system that is always up to date, and lists products and setup details, which many companies do not have yet. The interviewee stated: \emph{``If you consume machine-readable advisories, you actually need an asset management system so that you can manage the matching. And that is usually lacking on the customer side.''}

Overall, the interviews revealed that many of the problems with security advisories arise due to an inconsistent format. 
Nevertheless, automation is still viewed skeptically at present, as companies would have to adapt their current setup to automation.

\section{Main Study — Survey}
\label{sec:main-study}

\subsection{Study Design}
\paragraph{Questionnaire and Testing}

Similar to the interviews, the main part of the survey was structured around the three parts, 1) information gathering, 2) information processing and 3) decision-making, which we identified from the literature review. The content and wording of the questions was derived from the findings of the preliminary study.
The full questionnaire can be found in \Cref{sec:appendix-questionnaire-mainstudy}. The language of the survey was English, as we aimed to recruit as many participants as possible. First, the participants were asked whether they regularly work with security advisories. As we are focusing specifically on this target group, the survey was ended at this point for all participants who do not handle security advisories. 
The participants were then queried about their occupation and general IT security expertise, such as in which economic sector they are working, their profession and the size of their company. Next, we asked questions about how frequently security advisories are dealt with and familiarity with CSAF. This was followed by the first major part about information gathering, were we asked which types of channels the participants receive their advisories, such as email or messengers.
We then provided statements concerning the reception and sources of advisories, such as ``I receive security advisories that do not affect me'', which participants could agree or disagree with using a 5-point Likert scale. These statements were derived from the problems identified during the preliminary study.

Next came questions about the processing of security advisories, 
such as whether an asset management system or automation is used. This was also followed by statements, originating from the preliminary study, the participants could agree or disagree with, this time concerning the processing of security advisories, such as ``Automation is essential to handle security advisories efficiently''. An attention test was also included here. After this, the participants were asked questions concerning decision-making, e.g., if they are involved in this process and how important are some factors, such as company policies and available resources.

Finally, the participants were given an opportunity to comment in a free text field if something was not covered by the previous questions.
The survey was concluded with demographic questions.

We conducted five incremental test runs, each with four to five testers. After each test round, the collected feedback  was integrated.  The testers were mainly IT security researchers of our lab and institutes who research or work in the IT security context. For the last round, we recruited people from the industry who work with security advisories on a daily basis.

\paragraph{Ethics}

The study was approved by the data protection office of Friedrich-Alexander-Universität
Erlangen-Nürnberg (FAU). At the beginning of the survey, the participants were informed about their data protection rights and the purpose of the study in an informed consent form. Participants were only able to take part in the survey if they gave their consent. There was no compensation, participation was voluntary. In addition, participants could interrupt or cancel the survey at any time without any disadvantages. The demographic questions were optional. An ``Prefer not to say'' option was offered and preselected for all demographic questions. All responses were anonymized. The survey was carried out on a self-hosted LimeSurvey\footnote{\url{https://www.limesurvey.org}} instance at our university.

\paragraph{Data Analysis}
For the quantitative analysis, the data was first extracted from LimeSurvey as JSON files and then analyzed using Python scripts. For correlation analysis, $\chi^2$ was used as the variables are categorical, and the effect sizes were determined by Cramér's $V$ \citep{1999-cramer}.
Considering the qualitative analysis, we followed the exclusion criterion of \citet{2019-mcdonald} as ``coding requires little interpretation'' \citep[p. 72:3]{2019-mcdonald}. Since there were only 20 comments in total, and these were brief and clearly understandable, e.g., ``The quality of writing in a security advisory is important'' or ``I consider qualified automatic actions on reported threats to be completely impossible'', we decided not to code them. 
One researcher summarized the free text answers. They were then discussed by two researchers to decide on the final findings and used to illustrate the results in the following.

\paragraph{Recruitment}

In order to advertise the online survey for international participants, 
we first posted the survey on fitting Reddits\footnote{For example \url{https://www.reddit.com/r/AskNetsec}} and used personal contacts who spread the survey on LinkedIn. We also advertised the survey using IT security mailing lists such as Fulldisclosure\footnote{\url{https://seclists.org/fulldisclosure/}} and Blueteamsec newsletter\footnote{\url{https://bluepurple.binaryfirely.com}}. 
With the assistance of DFN-CERT, the link to the survey was distributed to
the DFN community.

\subsection{Participants}
\label{sec:participants}
The survey ran from September 2023 to February 2024 and 230 people completed it. Some participants were excluded as they did not agree to the privacy policy or did not pass the attention test, resulting in a total of 197 valid responses.

An overview over the participants' demographics is presented in \Cref{tab:demographics}. The participants were on average 44 years old (median 45 years). About 80\% identified as male, 3\% as female, 1\% as diverse and 16\% preferred not to disclose this information. 
About half of the participants come from Germany, 18\% come from another European country and 8\% from the USA. Regarding education, 48\% of participants have a Master's degree and around 24\% have a Bachelor's degree. Only about 12\% of the participants have no academic or similar degree.

\begin{table}
	\begin{center}
		\begin{tabularx}{\linewidth}{l|X X}
			\toprule
			& $N$ & $\%$ \\
			Total & 197 & 100,0 \\\hline
			Male & 158 & 80.2\\
			Female & 6 & 3.0\\
			Diverse & 2 & 1.0\\
			N/A & 31 & 15.7\\\hline
			18-29 years & 11 & 5.6\\
            30-39 years & 41 & 20.8\\
			40-49 years & 44 & 22.3\\
			50-60 years  & 37 & 18.8\\
			Above 60 years & 10 & 5.1\\
			N/A & 54 & 27.0\\\hline
			Germany & 103 & 52.3\\
			Other European country & 35 & 17.8\\
			USA & 16 & 8.1\\
			Other country & 8 & 4.1\\
			N/A & 31 & 15.7\\\hline
			No academic education & 24 & 12.2\\
			Bachelor's degree & 47 & 23.9\\
			Master's degree & 94 & 47.7\\
			Ph.D. & 9 & 4.6\\
			Other & 4 & 2.0\\
			N/A & 20 & 10.2\\\hline
			1-5 years work experience & 27 & 13.7\\
			6-10 years work experience & 26 & 13.2\\
			11-20 years work experience & 60 & 30.5\\
			More than 20 years work experience & 84 & 42.6\\\hline
			None or basic IT security knowledge & 1 & 0.5\\
			Intermediate IT security knowledge & 25 & 12.7\\
			Advanced or expert IT security knowledge & 171 & 86.8\\
			\bottomrule
		\end{tabularx}
		\vspace*{0.5em}
		\caption{Overview of the participants' demographics, work experience and self-assessed IT security knowledge.}
		\label{tab:demographics}
	\end{center}
\end{table}

Most of the participants work in the IT field and have a job title that is directly related to IT security, such as ``security analyst'', ``cyber security specialist'', ``pentester'' and others. Other participants described their job title as ``head of IT'', ``IT lead'' or ``coordinator'', which means that they work as a leader or supervisor of a team and are presumably responsible for making decisions. The remaining participants were mostly system administrators or developers. 

On average, participants had 19 years of professional experience (median 20) and most considered their expertise in IT security to be advanced (42\%) or expert (45\%).
Most participants work in the education sector (25\%) or  information and communication services (23\%). 
The size of the companies was fairly evenly distributed. About 10\% of the participants work in a company with less than 50 employees, 17\% in one with 50-249, 16\% in one with 250-999, 36\% in one with 1,000-9,999, and 23\% in companies with over 10,000 employees.
Over 90\% of participants engage with security advisories on at least weekly basis, 9.1\% receive them monthly. Around 85\% of participants forward security advisories at least sometimes to potentially affected end customers and industry partners, and about 93\% of participants are at least partially involved in the decision-making process. 

\subsection{Results}
\label{sec:results}

\paragraph{Information Gathering (RQ1, RQ2)}
\label{sec:results-information-gathering}

Only about 12\% of the participants use a single channel, most use about 2-3 channels. Receiving security advisories per mail, web platforms and via colleagues 
are most popular, while messengers are less common. Other channels that were stated as free text answers were RSS feeds and social media, such as Mastodon or X, but also blogs, podcasts or company internal software. On average, participants are subscribed to 12 different mailing lists or similar services. We consider using different mailing lists as one channel.

\begin{figure*}
	\begin{center}
        \includegraphics[scale=0.6]{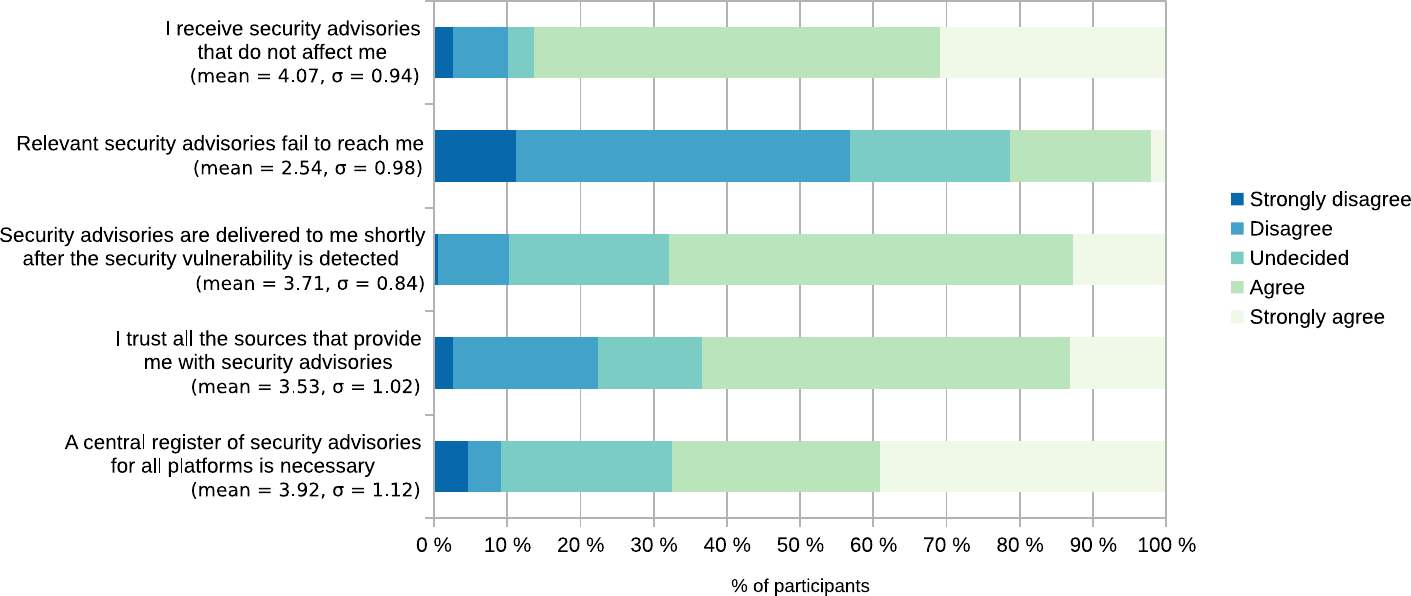}
	\end{center}
	\caption{Statements and participants' agreement for information gathering $(N = 197)$.}
	\label{fig:statements-information-gathering}
\end{figure*}

We presented a list of statements to the participants that they could agree or disagree with using a 5-point Likert scale. We coded the Likert scale from one (strongly disagree) to five (strongly agree) to calculate the mean and standard deviation, as recommended by \citet{2015-harpe}. The results are depicted in \Cref{fig:statements-information-gathering}.
Many participants frequently receive security advisories that do not affect them. 
In addition, the participants do not have the impression that they are missing out on important advisories. 
This is probably due to the fact that most participants obtain their information from several sources and have subscribed to multiple services. As a result, they receive more advisories that are not necessarily relevant to them, but do not miss the particularly important ones.
The advisories are sent to the participants shortly after they are published. 
The participants thus do not encounter problems with delayed information.
The question whether the participants trust their sources that provide them with security advisories received rather mixed opinions.
Although the majority agrees, there are also participants who do not trust all their sources. 
Many participants see the need for a central register for security advisories, but 25\% of the participants were undecided. P59 stated: \emph{``With multiple sources of security advisories available there is definitely a need for a standard format that is machine readable [...]'' }.

\paragraph{Information Processing (RQ1, RQ2)}
\label{sec:results-information-processing}

Around 60\% of participants use an asset management system to maintain their companies' or organizations' IT inventory. At least 20\% would like to introduce one. 

\begin{figure*}
    \begin{center}
		\includegraphics[scale=0.6]{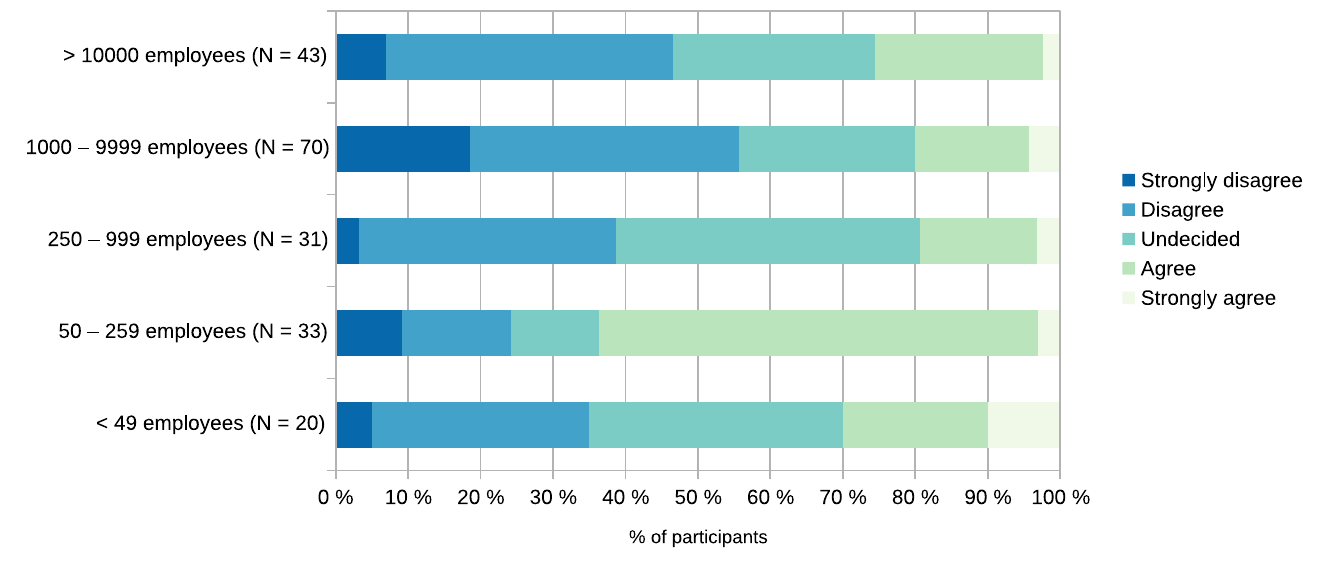}
		\caption{Agreement for ``It is easy to match the affected systems of a security advisory to our IT setup'' in relation to company size $(N = 197)$.}
		\label{fig:statements-information-processing-matching-companySize}
	\end{center}
\end{figure*}

\begin{figure*}
	\begin{center}
		\includegraphics[scale=0.6]{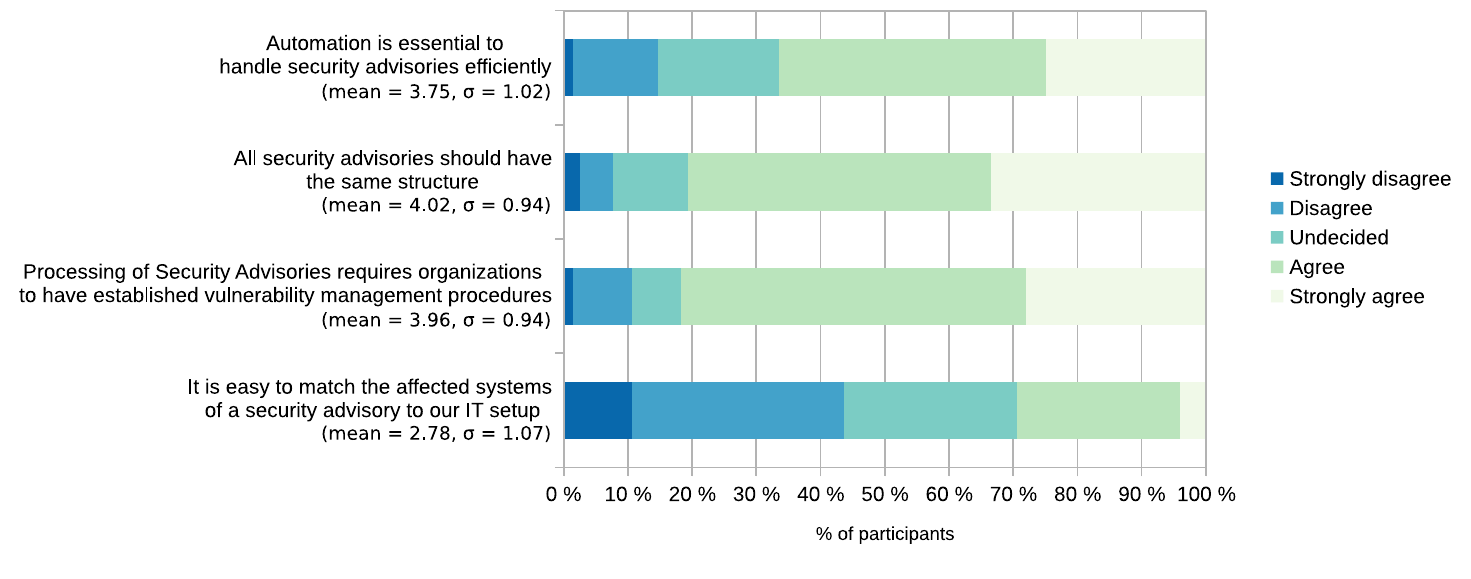}
	\end{center}
	\caption{Statements and participants' agreement for information processing $(N = 197)$.}
	\label{fig:statements-information-processing}
\end{figure*}

Around 26\% describe themselves as familiar with CSAF, while the majority are only slightly familiar (40\%) or not familiar at all (34\%).
This result is also reflected in the use of automation, as only about 33\% of participants already use automation and about 25\% have the desire to introduce it. However, around 42\% of participants do not want to use automation.
We dive deeper into usage of automation in \Cref{sec:results-automation}.
Considering the personal opinion about automation, most participants see automation as essential to handle security advisories in an efficient way (see \Cref{fig:statements-information-processing}).
They also clearly see the need for security advisories to have the same structure. 
The participants also agree that organizations need established vulnerability management procedures in order to process security advisories. 
In contrast, there is mixed agreement on how easy it is to match the affected systems listed in a security advisory to one's own IT setup. Around 24\% of participants think it is easy to identify the required information, while 43\% find it complicated and around 27\% were undecided. This confirms the finding from the qualitative study, in which the participants reported that the inconsistent format often made it very difficult to categorize the flood of affected systems and to identify whether one's own system is affected or not. 
When answers to this statement are compared with the size of the companies (see \Cref{fig:statements-information-processing-matching-companySize}), it is noticeable that participants who work in companies with 50-250 employees find matching easier than participants of larger and smaller companies. This could be because smaller companies often have a less complex IT infrastructure compared to larger ones, but have more resources to do matching than the smallest companies of up to 49 employees.

\begin{figure*}
	\begin{center}
		\includegraphics[scale=0.6]{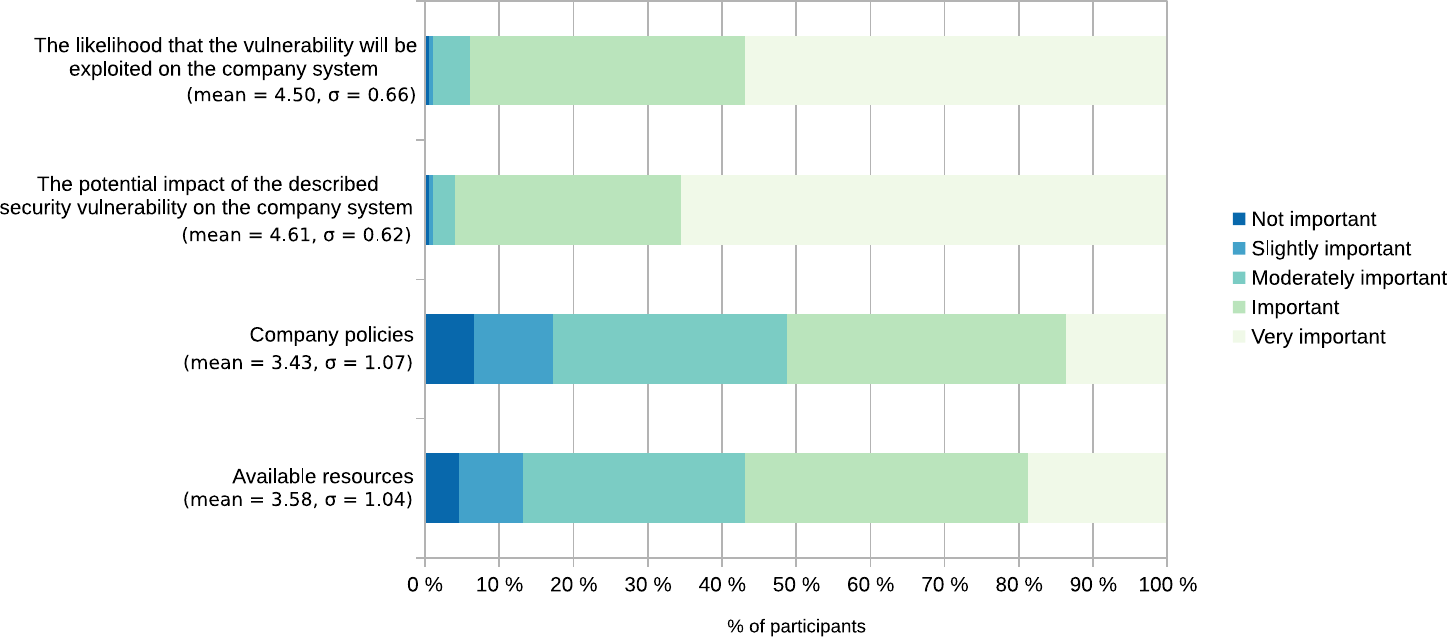}
	\end{center}
	\caption{Factors that may influence decision-making regarding the security issue described in an advisory, and their importance to the participants $(N = 197)$.}
	\label{fig:statements-decision-making}
\end{figure*}

\paragraph{Decision Making (RQ1, RQ2)}
\label{sec:results-decision-making}
After the important information has been extracted from the security advisory, a decision must be made on how to proceed with the vulnerability. Some factors play a more prominent role than others (see \Cref{fig:statements-decision-making}).
According to the participants, the most important factor is the likelihood that the vulnerability will be exploited and the potential impact of exploitation. The internal company policies and the available resources have a slightly smaller impact, although these are nevertheless considered to be important or very important by at least 50\% of the participants. 
If no decision can be made on the basis of the information available in security advisories, additional sources must be consulted. About 70\% of the participants use at least some additional sources to reach a decision. We asked these participants to describe their additional sources in more detail in a free text field. These sources are often public security scores that provide an approximate security assessment, such as CVSS, 
or databases from manufacturers and vendors that provide further information such as patch notes. Software used for vulnerability scanning or management is also utilized. Sometimes additional expertise is obtained from security consultants, peers or colleagues, or publicly available information is used, such as news sites, blogs or social media. In rare cases, internal company records of past incidents, decisions or best practices are also used.

\paragraph{Automation (RQ3)}
\label{sec:results-automation}

\begin{figure}
    \begin{center}
		\includegraphics[scale=0.5]{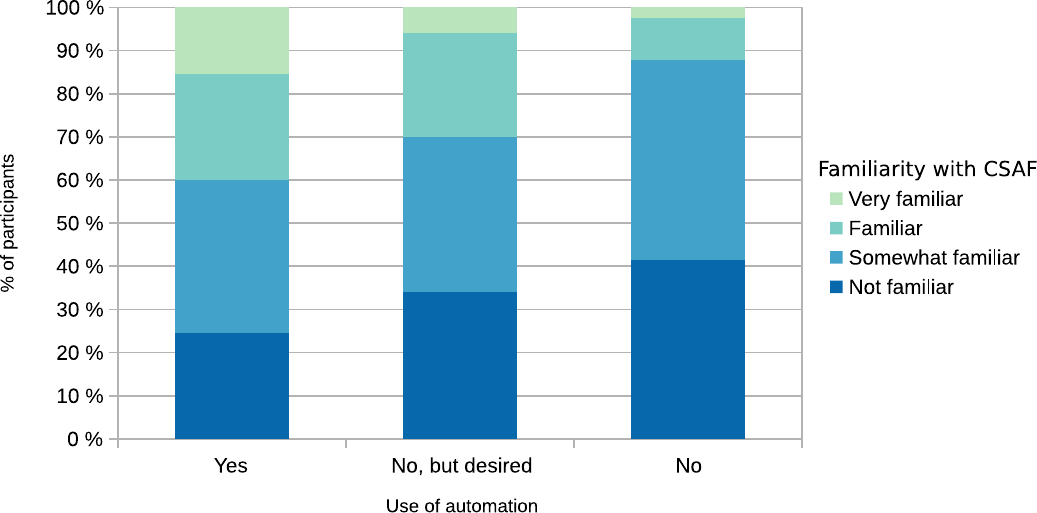}
		\caption{Use of automation in relation to CSAF familiarity $(N = 197)$; $\chi^2(6) = 18.19$, $p < 0.05$, Cramér's $V = 0.215$.}
		\label{fig:automation-csaf}
	\end{center}
\end{figure}

About 33\% of participants already use automation and about 25\% have the desire to introduce it. However, around 42\% of participants do not want to use automation. 
If the use of automation is correlated to familiarity with CSAF (see \Cref{fig:automation-csaf}), it seems that people who use automation are significantly more likely to know about CSAF ($\chi^2$ = 18.19, $p<0.05$).
Cramér's $V$ = 0.215, which can be interpreted as small to moderate effect size.

\begin{figure}
    \begin{center}
		\includegraphics[scale=0.5]{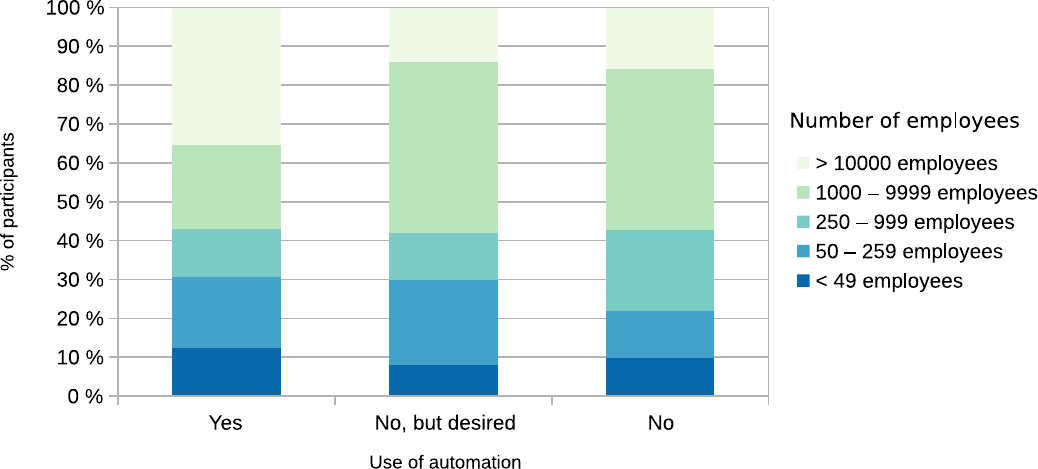}
		\caption{Use of automation in relation to company size $(N = 197)$; $\chi^2(6) = 18.34$, $p < 0.05$, Cramér's $V = 0.216$.}
		\label{fig:automation-company}
	\end{center}
\end{figure}

Another finding is that automation seems to be significantly related to the size of the company, see \Cref{fig:automation-company} ($\chi^2$ = 18.34, $p<0.05$). Cramér's $V$ = 0.216, which indicates a small to moderate effect size. 
Larger companies with more than 10,000 employees tend to use automation more than smaller companies. This is not surprising, as larger companies often have more resources.

\begin{figure}
    \begin{center}
		\includegraphics[scale=0.5]{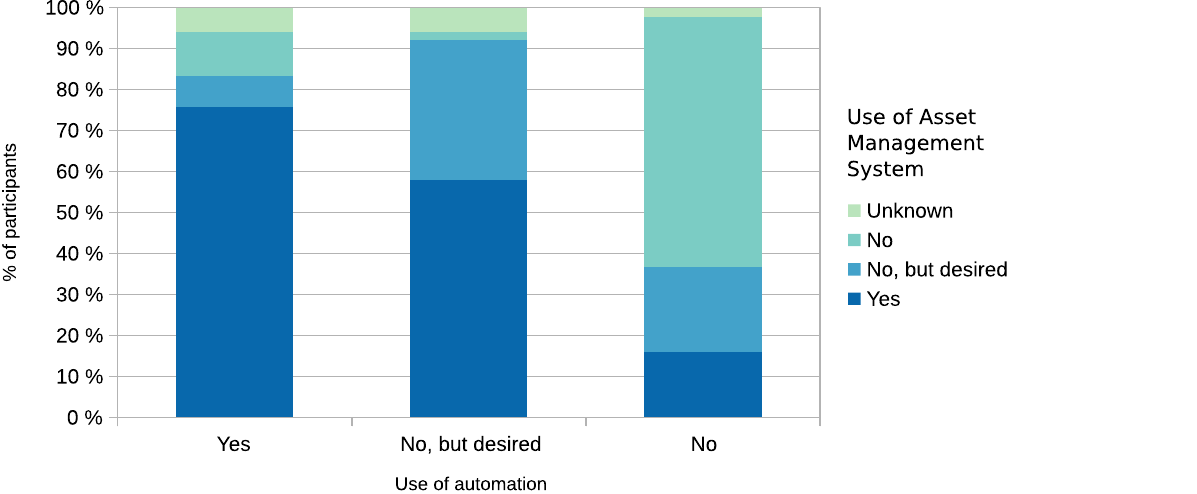}
		\caption{Use of automation in relation to usage of asset management systems $(N = 197)$; $\chi^2(6) = 18.56$, $p < 0.05$, Cramér's $V = 0.217$.}
		\label{fig:automation-assetManagement}
	\end{center}
\end{figure}

The use of asset management systems also appears to be significantly correlated 
to the use of automation ($\chi^2$ = 18.56, $p<0.05$), see \Cref{fig:automation-assetManagement}. The effect size ranges from small to moderate (Cramér's $V$ = 0.217). 
Most likely, the use of an asset management system simplifies the introduction of automation.

\paragraph{Additional remarks}
\label{sec:results-additional remarks}
At the end of the questionnaire, we gave the participants the opportunity to express further thoughts on the topic in the form of a free text field. Some participants again emphasized the poor quality of security advisories, which manifested itself in poor writing quality, bad wording or wrong product and version numbers. \emph{``Security advisories often contain wrong data,''} wrote P51, and P14 stated: \emph{``People must learn how to write.''} Other participants used this section to criticize existing security scores such as CVSS. Some participants also emphasized once again that they are skeptical about automation, as P137 described: \emph{``I consider qualified automatic actions on reported threats to be completely impossible''}. P97 explained this further: \emph{``A fully automated process to install security updates is not feasible for all used products. Especially propriatary [sic] software [...] have a worse testing in place before releasing a software update and might break a running system. In recent years this problem has worsened a lot as the vendors internal policy seems to reduce the spending on testing and internal reviews of their products.''} P131 stated: \emph{``Vulnerability management is such a complex challenge that it generally lacks ressources [sic].''}

\section{Discussion}
\label{sec:discussion}

\subsection{Current State of Advisories (RQ1, RQ2)}
The processing of security advisories can be divided into different phases, as it was identified using related work \citep{2019-li, 2020-tiefenau}. First, advisories must be received from sources. They must then be filtered according to relevance. Then a decision has to be made on how to proceed with the vulnerability.
However, little was known about how exactly these phases are structured and what role security advisories play in them which was investigated through this study.
The results showed that security advisories are processed on an almost daily basis which requires a high level of resources. In most cases, several sources are used simultaneously, as one source is often not sufficient to obtain all the information required. The participants therefore see a need for a central source that distributes the security advisories. Due to the different sources, many security advisories are also received that are not applicable for the users at all. In this case, users must first go through the time-consuming step of manually analyzing the security advisory, and determining whether their own product is affected which is very cumbersome due to the inconsistent format. In addition, the manufacturers often do not use the same unique product identifiers in the security advisory, but internal identifiers, which \citet{2020-farhang} also noted, though they did not focus on the user side. This makes this step unnecessarily resource-intensive and complicated.

We found evidence that the participants see the need for a standardized format with a uniform structure, as it would solve this problem.
The decision-making process then depends on the company's objectives, with the likelihood of exploitation and the impact of the vulnerability being particularly important according to the participants. Additional sources are often used in the decision-making process which in turn could be linked or integrated into security advisories in order to save additional resources.

\subsection{Automation (RQ3)}
Automation could save additional resources and simplify time-consuming processes, which is why automation is emphasized as particularly important \citep{2017-ramnani, 2008-fenz-semantic, 2008-fenz-fortification}. Nevertheless, some participants are not currently planning to switch to automation. This may be mainly due to the fact that some prerequisites must first be met for automation to be compatible. For example, an integrated asset management system must be in place, maintained, and managed to provide an up-to-date list of product numbers and current version numbers. Introducing such a tool into an existing system can involve a great deal of effort, for which current company resources may not be sufficient. Some participants are skeptical about automation, as they believe that automation cannot solve all problems, such as the problem of inconsistent product identifiers.

\subsection{Need for CSAF (RQ4)}
CSAF, as a standardized format, would solve some of the problems mentioned above. It is likely that users will continue to receive irrelevant advisories, as the problem of inconsistent product identifiers is in the hands of the manufacturers. However, with CSAF it would be easier to manually find out whether users' systems are affected, or this step could be even fully automated.
More information could also be added to security advisories, which the participants of this study would welcome, as they believe that there cannot be too much information in a security advisory. Nevertheless, many information entries in CSAF are optional, and it is up to the manufacturers and vendors to decide how much information they provide and how well written it is. The main criticism from participants is that some manufacturers pay too little attention to their advisories, and that the general quality of advisories is rather poor.

CSAF is the right way to achieve automated processing of security advisories and to simplify complex processes in order to save valuable resources. Nevertheless, there are some problems that even a standardized format cannot solve. If information fields are optional, it is up to the manufacturers and vendors to decide how carefully they enter this information, and what its quality is. In addition, an automated format only makes sense if users are prepared to accept automation and adapt their own systems to it which was not always the case in our study.

\subsection{Limitations}
\paragraph{Internal Validity} The survey aimed to investigate how people process security advisories, so we recruited individuals who regularly encounter security advisories using a filter question. We avoided mentioning CSAF in the beginning of the survey to include both familiar and unfamiliar participants. We conducted multiple rounds of testing with various participants, including four industry professionals, to ensure that the questionnaire was understandable and purposeful to our research questions.
Participants were mainly asked for facts, not opinions, to minimize potential question order effects. Although we couldn't confirm if participants took the survey more than once, we believe that this is unlikely, as we offered no compensation. 

\paragraph{External Validity} This study might be limited by the small sample size of the qualitative preliminary study, as only three people were interviewed. However, as the interviews served to gain an initial insight and collect problems that would then be examined in more detail in the survey, we did not expect that the interviews to be representative of the community who deals with security advisories regularly. 
Another limitation is that slightly more than 50\% of participants of the main survey are from Germany, though efforts were made to recruit internationally. With participants representing a geographically more diverse group, different conclusions might have been reached. We took the utmost care to find out important topics and issues concerning security advisories through literature review and three interviews. Nevertheless, this survey might have left out some aspects of working with security advisories if they did not come up in the preparatory work. 
Due to the survey's quantitative nature, only the frequency of identified aspects can be described, and empirical evidence of the reasons behind them cannot be provided.

\section{Conclusion}
\label{sec:conclusion}

In this work, we investigated how security advisories are currently used, where problems arise and whether CSAF can improve this process. 

First, we conducted a qualitative preliminary study in which we interviewed three participants about their handling of security advisories. A particular focus here was on the establishment of automation and the participants' opinions on this, as CSAF emphasizes automation. We then conducted a quantitative online survey with 197 participants. The results show that users encounter different problems when processing security advisories, and that these problems occur at different phases. For example, participants often receive many irrelevant security advisories and then have to filter them manually. This is particularly resource-intensive, as the version numbers concerned are often in a non-uniform format and confusing. CSAF could solve this problem with a standardized format. Many participants see the advantages of CSAF and automation, and recognize the need for improvements. Nevertheless, there are hurdles to overcome as some users are
not yet fully convinced by automation or lack the prerequisites for its use. 

As over half of the participants came from Germany, future work could extend the research to other countries.
Further studies could also explore manufacturers opinions on CSAF, including their process for detecting potential security flaws and creating security advisories.
Another follow-up study could evaluate the outcomes of this research. In this case, the results could be discussed with CSAF users to identify further trends. Alternatively, the findings could be discussed with the developers of CSAF in order to potentially contribute to improvements.

This work showed that CSAF is not yet widely used. If it becomes more widespread in the future, a study could explore whether CSAF was actually able to solve the problems identified in this study and improve the processing of security advisories.

\paragraph{Acknowledgements}
We thank Martin Waleczek, Tobias Dussa and Thomas Schreck for their valuable support in recruiting participants.
We also thank Freya Gassmann, who conducted the correlational statistical analysis.
We thank the testers and participants of the survey. 
We thank the anonymous reviewers who improved the paper with their feedback. The authors were supported by the German
Federal Ministry of Education and Research under grant 16KIS1271K.

\paragraph{Author Contribution Statements}
\textbf{Julia Wunder:} Conceptualization, Methodology, Formal analysis, Visualization, Supervision, Project administration, Writing - Original Draft (Chapters 1, 2.3, 2.4, 3, 4, 5, 6), Writing - Review \& Editing, 
\textbf{Janik Aurich:} Conceptualization, Methodology, Formal analysis, Investigation, Data Curation, Writing - Original Draft (Chapters 2.1, 2.2), Writing - Review \& Editing
\textbf{Zinaida Benenson:} Visualization, Supervision, Project administration, Writing - Review \& Editing, Funding acquisition

\bibliographystyle{ACM-Reference-Format}
\bibliography{csaf-paper}

\appendix

\section{Interview Guide of the Preliminary Study}
\label{sec:appendix-interview-guide}

\subsection{Introduction}
\begin{itemize}
    \item Greeting, short introduction of people performing the interview + taking notes
    \item  Short outline of thesis topic and why we’re conducting interviews in the first place
    \item  Notice on video / audio recording + approval by the interviewee
    \item  Thank interviewee for their time and help
\end{itemize}

\subsection{General demographic information}
\begin{itemize}
    \item Age, gender, country of residence, education
    \item Employment
    \begin{itemize}
        \item job title, employment form
        \item general routine / tasks
    \end{itemize}
\end{itemize}

\subsection{Security advisories general}
\begin{itemize}
    \item How often security advisories come up during work
    \item  General view on security advisories in their current state \emph{(positive, neutral, negative)}
\end{itemize}

\emph{Give an outlook of the following three categories and get across that they are to be analyzed one by one independent of each other.}

\subsection{Information gathering}
\emph{Explain that this is only about collecting information.}
\begin{itemize}
    \item How are security advisories received
    \begin{itemize}
        \item actively searched for
        \item notifications
    \end{itemize}
    \item Is there anything that makes gathering information particularly hard?
    \begin{itemize}
        \item volume
        \item sources
        \item availability
    \end{itemize}
    \item Is there anything you can think of that would improve the information gathering process?
    \begin{itemize}
        \item time
        \item security
    \end{itemize}
\end{itemize}

\subsection{Information processing}
\emph{Assume all the necessary information is now collected. Clarify that this is only about the presentation of information and not about comprehension.}
\begin{itemize}
    \item In general, how is information processed
    \begin{itemize}
        \item manual reading / skimming
        \item automated work
    \end{itemize}
    \item Are there any obstacles that make processing advisories difficult for you?
    \begin{itemize}
        \item language
        \item redundancy
        \item complexity
    \end{itemize}
    \item Is there anything you would wish for that would optimize information processing?
\end{itemize}

\subsection{Decision making}
\emph{So now the information is processed, what happens next.}
\begin{itemize}
    \item What factors are most influential in forming a decision on how to proceed?
    \begin{itemize}
        \item perceived severity
        \item available resources
    \end{itemize}
    \item Under what conditions is it impossible to utilize security advisories to make an informed decision?
    \begin{itemize}
        \item work environment / policies
        \item security advisory content
        \item general comprehension of issue
    \end{itemize}
    \item Is there anything you would change or add to security advisories that would simplify the decision-making process?
    \begin{itemize}
        \item risk score
        \item contact information
    \end{itemize}
\end{itemize}

\subsection{Optional}
\emph{Anything to add?}

\subsection{Conclusion}
\begin{itemize}
    \item Thank for the interview, ensure that their participation is of great value to the study
\end{itemize}

\section{Questionnaire of the Main Survey}
\label{sec:appendix-questionnaire-mainstudy}
This is the questionnaire of the main survey. In the beginning, the questionnaire informs the participant about the purpose of the study, which is followed by the participation consent.

\subsection{Preselection Question}
Security advisories are used to communicate a security vulnerability in a system. They are usually issued by the manufacturer of the affected system and contain detailed information about the vulnerability as well as how to further proceed to mitigate the risk of being attacked.

\begin{itemize}
    \item Do you regularly encounter security advisories as part of your work?
    \begin{itemize}
        \item Yes / No
    \end{itemize}
\end{itemize}

(If ''No`` is selected, the survey ends with a short explanation that we seek participants who regularly encounter security advisories as part of their work.)

\subsection{Occupation and general IT security}
\begin{itemize}
    \item In which economic sector are you currently working?
    \begin{itemize}
        \item Accommodation and food services / Administrative and support service activities / Agriculture, forestry and fishing / Construction / Distributive trade sector / Education / Electricity, gas, steam and air conditioning supply / Information and communication services / Manufacturing / Mining and quarrying / Professional, scientific and technical activities / Real estate activities / Repair of computers and personal and household goods / Services (except transport and storage) / Transportation and storage services / Water supply, sewage, waste management and remediation activities / Other: (free text)
    \end{itemize}
	\item What is your profession? (free text)
    \item How large is the company you work for?
    \begin{itemize}
        \item < 49 employees / 50 - 249 employees / 250 - 999 employees / 1000 - 9999 employees / > 10000 employees
    \end{itemize}
    \item How many years of work experience do you have?
    \item Rate your level of knowledge in IT security.
    \begin{itemize}
        \item None / Basic / Intermediate / Advanced / Expert
    \end{itemize}
    \item How frequently do you engage with security advisories?
    \begin{itemize}
        \item Daily / Weekly / Monthly / Yearly
    \end{itemize}
    \item Common Security Advisory Framework (CSAF) is a standardized format for disclosing security vulnerabilities. It is designed to enable efficient creation and handling of security advisories with a focus on automation. How familiar are you with CSAF?
    \begin{itemize}
        \item Not familiar / Somewhat familiar / Familiar / Very familiar
    \end{itemize}
\end{itemize}

\subsection{Sources for security advisories}
\begin{itemize}
    \item Please indicate the types of channels through which you receive security advisories. (multiple-choice)
    \begin{itemize}
        \item Email / Messenger (SMS, Whatsapp, \dots) / Colleagues / Web platform / Search Engine / Other: (free text)
    \end{itemize}
    \item How many different mailing lists or other services that provide security advisories are you subscribed to?
    \item Below is a list of statements about receiving security advisories. Please answer to what extent you agree with the following statements:\\
    (Strongly disagree, Disagree, Undecided, Agree, Strongly agree)
    \begin{itemize}
        \item I receive security advisories that do not affect me.
        \item Relevant security advisories fail to reach me.
        \item Security advisories are delivered to me shortly after the security vulnerability is detected.
        \item I trust all the sources that provide me with security advisories.
        \item A central register of security advisories for all platforms is necessary.
    \end{itemize}
\end{itemize}

\subsection{Processing of security advisories}
\begin{itemize}
    \item Asset Management Systems are tools that are used by organizations to efficiently track and maintain their IT inventory.
    Does your organization use an Asset Management System to keep track of the IT inventory?
    \begin{itemize}
        \item Yes / No / No, but desired / Unknown
    \end{itemize}
    \item Do you use automation for processing Security Advisories?
    \begin{itemize}
        \item Yes / No / No, but desired
    \end{itemize}
    \item Below is a list of statements about processing security advisories. Please answer to what extent you agree with the following statements:\\
    (Strongly disagree, Disagree, Undecided, Agree, Strongly agree)
    \begin{itemize}
        \item Processing of Security Advisories requires organizations to have established vulnerability management procedures.
        \item Automation is essential to handle security advisories efficiently.
        \item It is easy to match the affected systems of a security advisory to our IT setup.
        \item This is an attention test, please mark ``Strongly agree''.
        \item All security advisories should have the same structure.
    \end{itemize}
    \item Do you forward security advisories to potentially affected end customers and industry partners?
    \begin{itemize}
        \item Yes / Sometimes / No
    \end{itemize}
\end{itemize}

\subsection{Decision-making based on security advisories}
\begin{itemize}
    \item After reviewing a security advisory, a decision has to be made on how to deal with the described security vulnerability. Are you involved in the decision-making process?
    \begin{itemize}
        \item Yes / Partially / No
    \end{itemize}
    \item Below is a list of factors that may influence decision-making regarding the security issues described in an advisory. Please indicate their importance based on your views.\\
    (Not important, Slightly important, Moderately important, Important, Very important)
    \begin{itemize}
        \item The likelihood that the vulnerability will be exploited on the company system.
        \item The potential impact of the described security vulnerability on the company system.
        \item Company policies
        \item Available resources
    \end{itemize}
    \item Which other factors — if any — influence the decision-making process as you see it? (free text)
    \item Do you use additional sources to reach a decision on how to progress?
    \begin{itemize}
        \item Yes / Partially / No
    \end{itemize}
    \item Which additional sources do you use? (free text)
\end{itemize}

\subsection{Additional Remarks}
\begin{itemize}
    \item Are there any additional remarks you would like to add to any of the aforementioned questions? (free text)
\end{itemize}

\subsection{Demographic information}
In the last section we would like to capture some demographic information about you, which will
enable us to analyze the collected data further.
\begin{itemize}
    \item What is your year of birth?
    \item What gender do you identify with?
    \begin{itemize}
		\item Male / Female / Diverse / Prefer not to say
	\end{itemize}
    \item What is your country of residence?
    \item What is your highest education level?
    \begin{itemize}
        \item Less than high school (no university/college entrance certificate) / Entrance certificate for university/college (high school diploma, GED, GCE, etc.) / Some college, associate degree or equivalent / Bachelor's degree or equivalent / Master's degree or equivalent / Professional degree (M.D., J.D., etc.), Ph.D. (doctoral degree) or equivalent / Prefer not to say / Other: (free text)
    \end{itemize}
\end{itemize}

\end{document}